\documentclass[twocolumn,showpacs,preprintnumbers,amsmath,amssymb,prl,superscriptaddress]{revtex4}

\usepackage{graphicx}
\usepackage{dcolumn}
\usepackage{bm}

\begin{document}

\title{Monte Carlo simulation of pressure-induced phase transitions in spin-crossover materials}

\author{Yusuk\'{e}~Konishi}
\affiliation{Department of Physics, The University of Tokyo,
Hongo 7-3-1, Bunkyo-ku, Tokyo 113-8656, Japan}
\author{Hiroko~Tokoro}
\affiliation{Department of Physics, The University of Tokyo,
Hongo 7-3-1, Bunkyo-ku, Tokyo 113-8656, Japan}
\affiliation{JSPS, 5-3-1 Koji-machi, Chiyoda-ku, Tokyo 102-0083, Japan}
\author{Masamichi~Nishino}
\affiliation{Computational Materials Science Center, National Institute
for Materials Science, Tsukuba 305-0047, Japan}
\affiliation{CREST, JST, 4-1-8 Honcho Kawaguchi, Saitama 332-0012, Japan}
\author{Seiji~Miyashita}
\affiliation{Department of Physics, The University of Tokyo,
Hongo 7-3-1, Bunkyo-ku, Tokyo 113-8656, Japan}
\affiliation{CREST, JST, 4-1-8 Honcho Kawaguchi, Saitama 332-0012, Japan}

\date{\today}

\begin{abstract}
Pressure-induced phase transitions of spin-crossover materials were 
simulated by a Monte Carlo simulation in the constant pressure 
ensemble for the first time. 
Here, as the origin of the cooperative interaction, we adopt 
elastic interaction among the distortions of the lattice due to the
difference of the molecular sizes in different spin states, i.e., the high 
spin (HS) state and the low spin (LS) state. 
We studied how the temperature dependence of the ordering process 
changes with the pressure, and we obtained a standard sequence of 
temperature dependences that has been found in changing other parameters 
such as strength of the ligand field 
(S.~Miyashita et al., Prog. Theor. Phys. \textbf{114}, 719 (2005)). 
Various effects of pressure on the spin-crossover ordering process are
examined from a unified point of view. 
\end{abstract}


\maketitle

Several spin-crossover (SC) compounds have been extensively investigated 
\cite{sc:6,sc:94,sc:60,sc:25,sc:26,sc:28,sc:98,sc:96,sc:97,sc:95}, 
and various theoretical analyses of the 
SC transitions have been reported \cite{sc:62,sc:63,sc:14,sc:20,sc:16,sc:77,sc:2,sc:1,sc:21,sc:89}.
In the SC compounds, a metal ion can be in either a low-spin (LS) or 
high-spin (HS) state, depending on the strength of the ligand field. 
Control of the spin state of SC compounds has been realized by applying 
external stimuli such as temperature, light-irradiation \cite{sc:6,sc:94,sc:60,sc:25,sc:96,sc:97,sc:98,sc:26,sc:20,sc:77,sc:28}, 
magnetic field \cite{sc:78,sc:66,sc:68,sc:70}, and pressure \cite{sc:99,sc:11,sc:12,sc:13,sc:100,sc:101}. 
It has been pointed out that cooperative interactions play an important role 
for the SC transitions. 
With such interactions, various types of SC transition 
are realized, e.g., a smooth crossover or a discontinuous first-order phase 
transition. 
For modeling of the interaction mechanism, 
a model with an Ising-type short-range interaction, e.g., 
the Wajnflasz-Pick (WP) model, 
has been proposed, and various aspects of cooperative behavior have been
successfully explained \cite{sc:2,sc:1,sc:20,sc:62}.
As the ligand field is changed, SC transitions show a sequence
of temperature dependences of HS fraction $f_{\rm HS}(T)$; 
(I) a smooth transition, (II) hysteresis, 
(III) hysteresis with a low-temperature metastable HS phase, and (IV) 
a HS phase stable at all temperatures.
We found that this sequence also appears with changing degeneracy  
or strength of the interaction \cite{sc:2,sc:1}.
Thus we call this the {\it generic sequence}.
However, the origin of the interaction was not clear. 
Recently, it has been pointed out that the elastic interaction between 
distortions of the lattice due to the molecular size difference between 
the HS and LS states can induce phase transitions of the spin state 
\cite{sc:21}.
In order to study the elastic interaction, besides the spin state,
the positions of the molecules must be treated as dynamical variables 
to be equilibrated. 
This degree of freedom causes a change of the system volume.
Therefore, the pressure $P$ becomes an important parameter for 
describing the system.
Thus, we are now at a stage where 
we can study the pressure effect by direct numerical study.
In a previous study \cite{sc:21}, we demonstrated that the  
elastic interaction can cause a spin transition in a 2D system with 
an open boundary condition. 
In this study, we adopt a similar elastic model in 3D. 
Although intramolecular potentials were taken into account besides 
intermolecular potentials in Ref.~\cite{sc:21}, in the present study we 
take into account two levels (LS and HS) with two different molecular 
sizes as the molecular state for simplicity.
Here we adopt a method for the constant pressure ensemble in 
a periodic boundary condition.
Moreover, we adopt a Monte Carlo (MC) method where we can easily control the 
degeneracy of the spin state. 
By this method, we succeeded for the first time in demonstrating 
pressure-induced phase transitions in a spin-crossover system.

The pressure effect has been one of the most important 
characteristics observed in SC compounds, and various interesting properties
have been observed, e.g., a 
shift of the transition temperature, increase or decrease of hysteresis, 
stabilization of the LS state for the whole temperature region, 
and so on \cite{sc:99,sc:11,sc:12,sc:13,sc:100,sc:101}.
To understand the pressure effects, a theoretical analysis based on 
each experimental phenomenon and the comprehensive theory of the 
free energy of the mean-field model has been reported \cite{sc:63,sc:14,sc:16}. 
In the present work, we study these effects by a direct numerical 
method, adopting a microscopic Hamiltonian and demonstrating 
the fundamental aspects in a unified 
microscopic picture.

\begin{figure}[tb]
\includegraphics[keepaspectratio=true,width=70mm]{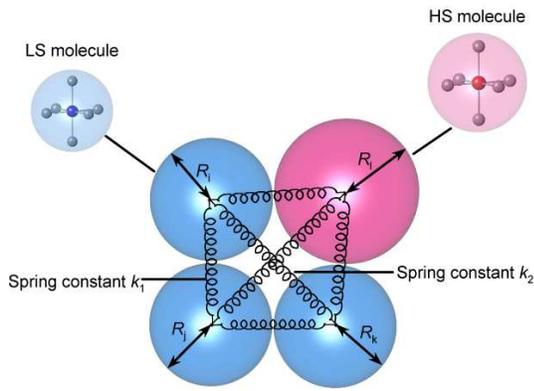}
\caption{
(Color online) Schematic illustration of the present model. 
HS/LS molecule consists of Fe atom (red/blue circle) and 
six ligands (gray circles). 
}
\label{fig-ill}
\end{figure}

We performed MC simulation on the simple cubic lattice.
In the spin-crossover materials, the molecule at a lattice point 
may be in the HS or LS state. 
We express the spin state at the $i$-th site by $n_i$, 
which equals 0 for the LS state and 1 for the HS state.
As important ingredients of the spin-crossover material, 
we set the energy difference of the state $D(>0)$, 
and the degeneracies of the states: $g_{\rm HS}$ and 
$g_{\rm LS}$ for the HS and LS states, respectively.
These properties are represented by the on-site Hamiltonian
\begin{equation}
{H}_0=D\sum_in_i,
\end{equation}
In order to introduce interaction between spins, an Ising-like interaction
$-J\sum_{<ij>}n_in_j$ was adopted in the WP model.
Instead, we have attributed the interaction to 
the elastic interaction between lattice distortions caused by the difference of
the molecular size between the HS or LS states. 
Therefore, we introduce the elastic interaction between the molecules:
\begin{eqnarray}
H=H_{\rm nn}+H_{\rm nnn} + H_0, \\
H_{\rm nn}=\frac{k_1}{2}\sum_{\langle i,j \rangle}[r_{ij}-(R_i+R_j)]^2, \\
H_{\rm nnn}=\frac{k_2}{2}
\sum_{\langle\langle i,j \rangle\rangle}[r_{ij}-\sqrt{2}(R_i+R_j)]^2,
\end{eqnarray}
where $r_{ij}$ is the distance between the molecule on the 
$i$-th and $j$-th sites, 
and $k_1$ is the corresponding spring constant (Fig.~\ref{fig-ill}). 
$H_{\rm nn}$ expresses elastic interaction between 
nearest-neighbor pairs ($\langle i,j\rangle$).
The interaction is a function of the molecular radius 
$R_i(= (1-n_i)R_{\rm LS}+n_iR_{\rm HS})$ of the $i$-th site, where $R_{\rm HS}$ 
and $R_{\rm LS}$ are the molecular radius of HS and LS states, respectively. 
We set the ratio of the radii to be $R_{\rm HS}/R_{\rm LS}=1.1$. 
$H_{\rm nnn}$ 
expresses elastic interaction 
for next-nearest neighbor pairs ($\langle\langle i,j\rangle\rangle$).
In this study, we set the ratio of the spring constants, $k_1/k_2$, 
to be 10 \cite{spring}.

For the simulation, we adopt the $NPT$-MC method 
\cite{sc:18} for the
isothermal-isobaric ensemble with the number of molecules $N$, the pressure of 
the system $P$, and the temperature $T$. 
The thermodynamic potential for the isothermal-isobaric ensemble is the enthalpy,
\begin{equation}
H=E+PV,
\end{equation}
where $E$ is the energy given by Eq.~(1) and $V$ is the volume of the system.
The states of the system are specified by $4N+1$ variables 
$(n_1,\cdots,n_N,{\bf r}_1,\cdots,{\bf r}_N,V)$. 
In the $NPT$-MC method, we have the following balance condition for
the transition probabilities $\{w_{i\to k}\}$: 
\begin{equation}
\frac{w_{i\to k}}{w_{k\to i}}=\exp(-\beta\Delta W), 
\label{detailed-balance}
\end{equation}
where
\begin{equation}
\Delta W=(E_k-E_i)+P(V_k-V_i)-NT\log(\frac{V_k}{V_i}).
\label{deltaw}
\end{equation}

The scheme of the simulation is as follows: 
(i) Choose a molecule randomly.
(ii) Choose a candidate spin state $n_i=0$ or 1 by the probability
$g_{\rm LS}/(g_{\rm LS}+g_{\rm HS})$ or $g_{\rm HS}/(g_{\rm LS}+g_{\rm HS})$, 
respectively. 
(iii) Choose a candidate position of the molecule (${\bf r}_i$).
Here, we use the scaled coordination length ${\bf s}={\bf r}_i/L$ ($L=V^{1/3}$).
We choose the candidate position as
\begin{eqnarray}
s_x\to s_x+\delta\xi_x,\quad s_y\to s_y+\delta\xi_y,\quad 
s_z\to s_z+\delta\xi_z,
\end{eqnarray}
where $\xi_x,\xi_y$ and $\xi_z$ are random numbers between $-1$ and 1, 
and $\delta$ 0.005. 
(iv) Update the state by the Metropolis method.
(v) Repeat the above update $N$ times 
(vi) Choose a candidate for a new size of the system with a random number $\xi_L$
\begin{equation}
L\to L+\lambda\xi_L.
\end{equation}
(vii) Update the size $L$.
Here, $\lambda$ is taken equal to $0.08N^{1/3}R_{\rm LS}$.
In this study, we performed 10000 MCSs for transient steps and 10000 MCSs 
to measure the physical quantities. 
The system size is $N=16\times16\times16$, which is enough to study 
thermal properties.

\begin{figure}[tb]
\includegraphics[keepaspectratio=true,width=70mm]{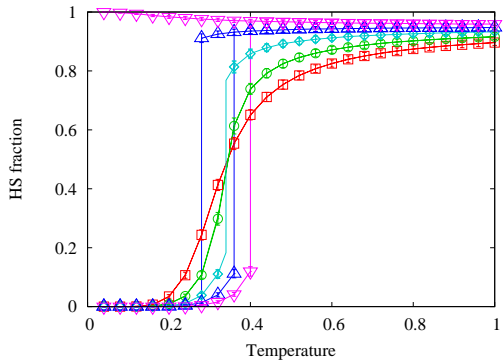}
\caption{
(Color online) Temperature dependence of the HS fraction $f_{\rm HS}(T)$ 
with $g = 20, D=1,P=0.01$, 
and $k_1=10$ (red squares), 20 (green circles), 30 (light blue diamonds), 
40 (blue triangles), and 50 (pink inverted triangles).
Solid lines are guides for the eye.
}
\label{fig-a}
\end{figure}
\begin{figure}[tb]
\includegraphics[keepaspectratio=true,width=75mm]{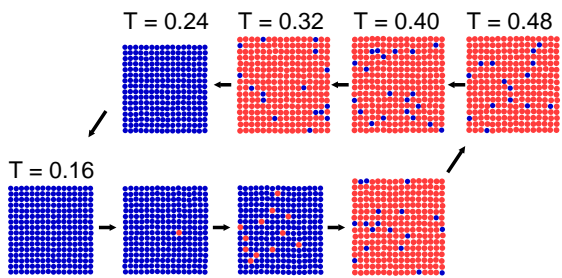}
\caption{
(Color online) Two-dimensional snapshots at various temperatures with 
$g = 20, D = 1, P = 0.01$, and $k_1=40$. 
Gray (red online) and black (blue online) circles are HS and LS molecules, respectively.
}
\label{ss}
\end{figure}

First, we study how the types of temperature dependence 
of the HS fraction $f_{\rm HS}(T)$ change with the spring constant $k_1 (=10k_2)$.
In Fig.~\ref{fig-a} we depict $f_{\rm HS}(T)$ for various values of $k_1$ 
with $g(=g_{\rm HS}/g_{\rm LS})=20,D=1,P=0.01,$ and $k1=10,20,30,40,$ and 50. 
When $k_1$ is small, e.g., $k_1=10$, the transition is gradual and 
hysteresis is not observed. 
As $k_1$ becomes large, the transition becomes sharp. 
For $k_1=40$, hysteresis is observed between cooling and warming processes. 
Here, the transition temperature is 0.28 for the cooling process 
and 0.36 for the warming process. Snapshots of the spin configuration 
(a two-dimensional section of the lattice) are shown in Fig.~\ref{ss}. 
When $k_1$ becomes larger, i.e., $k_1=50$, the transition 
does not take place in the cooling process and the HS phase is maintained 
down to $T=0$. 
This observation indicates the existence of a HS metastable phase 
at low temperatures \cite{sc:2,sc:1}.
In the warming process from $T=0$, the LS phase is transformed to the HS phase 
at $T=0.4$.

\begin{figure}[tb]
\includegraphics[keepaspectratio=true,width=70mm]{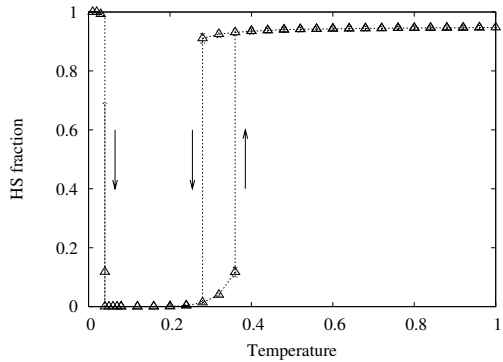}
\caption{
Temperature dependence of the HS fraction with $g=20,D=1,P=0.01$, 
and $k_1=40$.
Dotted line is a guide for the eye.
}
\label{fig-b}
\end{figure}

We find that the change of $k_1$ causes a sequence of $f_{\rm HS}(T)$ 
that agrees with the {\it generic sequence} proposed in our previous papers.
Thus, we expect that there is a case where 
a low-temperature HS metastable phase and thermal hysteresis are observed,
which was found characteristic of this type of ordering processes and
also was experimentally confirmed \cite{sc:2,sc:1,sc:79}. 
In order to check the existence of the low-temperature HS metastable phase,
we studied the warming-up process from HS from $T=0$ 
for the system with $k_1=40$.
The temperature dependence is depicted in Fig.~\ref{fig-b}, where   
the HS metastable phase exists and relaxes to the LS phase at $T=0.04$. 
The LS phase changes to the HS phase at $T=0.36$ as we saw in Fig.~\ref{fig-a}.
Therefore, we find that the HS metastable phase and hysteresis are 
both observed. 
Now, we confirm that the present model realizes the {\it generic sequence} of 
$f_{\rm HS}(T)$ of the SC transitions. 

\begin{figure}
\includegraphics[keepaspectratio=true,width=70mm]{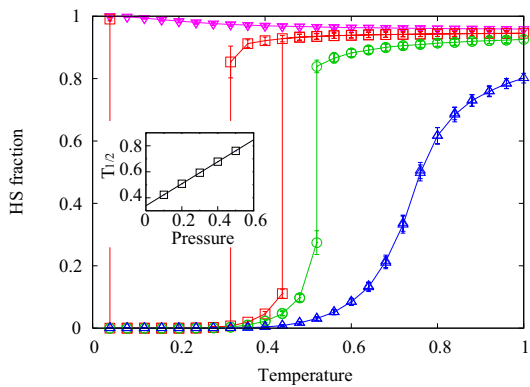}
\caption{
(Color online) Temperature dependence of HS fraction with $g=20,D=1,k_1=50$, 
and $P=0.01$ (pink inverted triangles), 0.1 (red squares), 
0.2 (green circles), 0.5 (blue triangles). 
Inset is the pressure dependence of $T_{1/2}$.
Solid lines are guides for the eye.
}
\label{fig-d}
\end{figure}

Next, we study the pressure effect on the SC transitions.
We study how $f_{\rm HS}(T)$ for $k_1 = 50$ changes
when the pressure increases. 
In Fig.~\ref{fig-d}, 
we depict $f_{\rm HS}(T)$ during the warming process from the HS state.
In the case of low pressure, $P=0.01$, the transition is not observed, 
as shown in Fig.~\ref{fig-a}.
When the pressure becomes $P=0.1$, the initial HS state 
relaxes to the LS phase at $T=0.04$, 
which indicates a low-temperature metastable HS state. 
Then, we find hysteresis with a jump at $T=0.32$ in the cooling process 
and at $T=0.44$ in the warming process.  
For $P=0.2$, the initial HS state immediately relaxes to the LS phase, 
which indicates no low-temperature metastable HS state. 
Here, the hysteresis disappears and $f_{\rm HS}(T)$ is shifted to 
the high-temperature side. 
When the pressure increases further high, $P=0.5$, 
the transition becomes gradual, and 
$f_{\rm HS}(T)$ is shifted to the high-temperature side.
We define $T_{1/2}$ at which $f_{\rm HS}=1/2$.
In the hysteresis region we define it as $(1/2-f_{\rm HS})$ in the 
warming process is equal the $(f_{\rm HS}-1/2)$ in the cooling process.
The pressure dependence of $T_{1/2}$  
is depicted in the inset of Fig.~\ref{fig-d}, which
indicates that $T_{1/2}$ increases linearly with the pressure. 

\begin{figure}[tb]
\includegraphics[keepaspectratio=true,width=70mm]{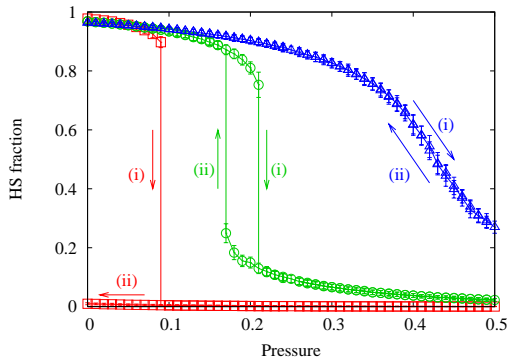}
\caption{
(Color online) Pressure dependence of the HS fraction with 
$g = 20, D = 1, k_1 = 50$, and 
$T = 0.3$ (red squares), 0.5 (green circles), and 0.7 (blue triangles). 
The numbered arrows indicate the pressure sequence. 
Solid lines are guides for the eye.
}
\label{fig-e}
\end{figure}

We also study the pressure dependences of $f_{\rm HS}(T)$ 
at various temperatures. 
In Fig.~\ref{fig-e}, $f_{\rm HS}(T)$ at $T =0.3,0.5$, and 0.7
for $k_1=50$
are depicted as functions of the pressure. 
Here, the HS phase is set as the initial phase. 
At $T = 0.3$, a small pressure $P=0.1$ induces the transition 
from the HS to the LS phases (red (i) arrow). 
The pressure-induced LS phase does not return to the HS phase in the process of
reducing pressure (red (ii) arrow). 
This irreversible pressure effect indicates that the pressure stabilizes 
the LS phase and destabilizes the HS phase. 
At $T=0.5$, the transition from the HS to the LS phase is observed at 
$P=0.22$ during the pressure-increasing process (green (i) arrow). 
During the pressure-reducing process, the transition from the LS to the 
HS phases is observed at $P=0.16$ (green (ii) arrow). 
That is, in this case 
we observe pressure-induced hysteresis.
At $T=0.7$, the transition between the HS and LS phases is smooth, and
the hysteresis disappears. 
The present observations indicate that the pressure plays a similar role to 
that of the temperature for the SC transition.

The $NPT$-MC method for a three-dimensional system was
established, and the effect of the pressure on the SC transition was studied.
To our knowledge, this is the first attempt to study the pressure effect
by direct numerical simulation, 
considering the local lattice distortions cased by the molecular 
size difference between HS and LS states in SC complex.
In particular,
we succeeded in observing a sequence of $f_{\rm HS}(T)$ as a function of the
pressure which agrees with that proposed as a general sequence the SC transitions.

In the present study, we kept the parameters $D$ and $g$ constant.
If we take into account this pressure dependence,
we can have a great variety of pressure dependences, which correspond to
complicated dependences observed in experiments.
For the next stage, we will study various pressure 
effects from the viewpoint of the present model and 
attempt to obtain a systematic understanding of the variety of  
pressure effects on the SC transitions.

The authors thank Professors Kamel Boukheddaden and 
Per Arne Rikvold for their valuable discussions.
This work was partially supported by Grant-in-Aid for Scientific
Research on Priority Areas
"Physics of new quantum phases in superclean materials" (Grant No.
17071011) from MEXT,
and also by the Next Generation Super Computer Project, Nanoscience
Program from MEXT.
This work was also partially supported by the MST Foundation.
The authors thank the Supercomputer Center, Institute for Solid State
Physics, University of Tokyo for the use of the facilities.


\begin{thebibliography}{33}
\bibitem{spring} The next-nearest neignbor interaction is introduced to 
maintain the cubic lattice, and the strength of $k_2$ is not important as 
long as the global shape of the lattice would not change.
\end{thebibliography}
\end{document}